\documentstyle[prl,aps,multicol,epsfig]{revtex}

\def\bege{\begin{equation}}
\def\ende{\end{equation}}
\def\bega{\begin{eqnarray}}
\def\enda{\end{eqnarray}}
\def\began{\begin{eqnarray*}}
\def\endan{\end{eqnarray*}}

\newcommand{\const}{\mbox{const}}

\newcommand{\equ}{Eq.}
\newcommand{\equs}{Eqs.}
\newcommand{\fig}{Fig.}
\newcommand{\figs}{Figs.}
\newcommand{\rem}[1]{}

\newtheorem{platz}{{Fig.}} % a dummy theorem to count figures!

\newcommand{\FIGo}[3]{\begin{figure}%
#3%
\caption[]{\footnotesize #2}%
\label{#1}%
\end{figure}}
\newcommand{\printfigcap}[2]{%
\rule{0cm}{1cm}%
Figure~\ref{#1}: 
#2 \\%
}

\rem{
\newcommand{\bstr}{1.48}  % double-spaced
\newcommand{\FIGo}[3]{%
\marginpar{\begin{platz} \label{#1} ~ \end{platz} \vspace*{1.5ex} }
}
\newcommand{\showfig}[1]{\begin{figure} #1
  \vbox{\vspace*{2.0cm}} \caption{Jan Wiersig} \end{figure} \clearpage }
}

\begin{document}

%for submission only
\rem{
{\large\noindent\vspace*{3ex}\\[3ex]
Jan Wiersig\\[1.5ex]
Max-Planck-Institut f\"ur Physik komplexer Systeme\\[1ex]
N\"othnitzer Str. 38\\[1ex]
D-01187 Dresden, Germany\\[1ex]
\begin{tabbing}
phone: \= \kill 
phone: \> +49 351 871 1223\\
fax:   \> +49 351 871 1999\\
email: \> jwiersig@mpipks-dresden.mpg.de\\
\end{tabbing}
}
\clearpage
\setlength{\footskip}{1cm}
\renewcommand{\baselinestretch}{\bstr} \normalsize
%end rem
}

\title{A pseudointegrable Andreev billiard}    
\author{Jan Wiersig}
\address{Max-Planck-Institut f\"ur Physik komplexer Systeme, D-01187
Dresden, Germany\\
jwiersig@mpipks-dresden.mpg.de}
\date{\today}
\maketitle
\begin{abstract} 
A circular Andreev billiard in a uniform magnetic field is studied. 
It is demonstrated that the classical dynamics is pseudointegrable in the same
sense as for rational polygonal billiards. The relation to a specific
polygon, the asymmetric barrier billiard, is discussed. 
Numerical evidence is presented indicating that the Poincar\'e map is
typically weak mixing on the invariant sets.  
This link between these different classes of dynamical systems throws some
light on the proximity effect in chaotic Andreev billiards. 
\\
\\
PACS numbers: 05.45.-a, 74.50.+r
\end{abstract}

%only for private preprint
\begin{multicols}{2}

\section{Introduction}
\label{sec:intro}
Billiards have played a prominent role in the understanding of classical and
quantum mechanics.   
In such a system, a particle moves freely in a domain with specular reflections
at the boundary (the angle of reflection equals the angle of
incidence); see, e.g.,~\cite{Berry81}.
Billiards of a different kind are realized in ballistic mesoscopic
samples connected to a superconductor~\cite{Andreev64,Andreev66}. The boundary
of such an {\it Andreev billiard}~\cite{KMG95} consists of normal-conducting
regions with specular reflections and superconducting regions with Andreev
reflections, whereby electronlike quasiparticles with charge $-q$, mass
$m$, and energy $\varepsilon$ above the Fermi energy in the normal
metal are retroreflected as holelike quasiparticles with charge $q$, mass
$-m$, and energy $-\varepsilon$. In the absence of a magnetic
field, retroreflected orbits are self-retracing and therefore periodic. The
presence of a magnetic field allows for a richer spectrum of dynamical
behavior~\cite{KMG95}.     

In this paper, we study an interesting Andreev billiard, the
circular Andreev billiard in a uniform magnetic field. Our analysis will
lead one to the conclusion that boundary points separating normal
and superconducting regions, henceforth called {\it critical points}, have
the same consequences on the classical dynamics as {\it critical corners} 
in rational polygonal billiards.  
In such a polygon, all angles $\alpha_j = m_j\pi/n_j$ between sides are
rationally related to $\pi$, where $m_j, n_j > 0$ are relatively prime
integers.   
The free motion inside a rational polygon is characterized as
{\it pseudointegrable}~\cite{RichensBerry81} since it shares some properties
of integrable systems:
(i) the phase space is foliated by two-dimensional invariant
surfaces~\cite{Hobson75,ZemlyakovKatok75};   
(ii) the flow on these surfaces is ergodic and not mixing~\cite{Gutkin96}, and,
in particular, not chaotic (see, e.g.,~\cite{AA68} for the definition of ergodic properties). 
Yet, in the presence of critical corners with $m_j > 1$ the dynamics is more
complex: 
(i) the genus of the surfaces is greater than one~\cite{RichensBerry81};
(ii) the dynamics is not quasiperiodic; and (iii) presumably weak mixing,
but this is proven only for a special subclass; see,
e.g.,~\cite{Gutkin96}. Numerical evidence for weak 
mixing has been reported in~\cite{ACG97,AGR00}.     

The paper is organized as follows. In Sec.~\ref{sec:andreev}, we
introduce the circular Andreev billiard and derive its Poincar\'e map,
which describes the collision-to-collision discrete dynamics.
We show that the dynamics is pseudointegrable, in the same sense as for
rational polygons.   
The relation to the asymmetric barrier billiard is discussed in
Sec.~\ref{sec:barrier}.  
Section~\ref{sec:dynamics} presents numerical evidence that the Poincar\'e
map is typically weak mixing on the invariant sets. 
In Sec.~\ref{sec:con}, we draw conclusions and give an outlook.

\section{The circular Andreev billiard}
\label{sec:andreev}
Let us first consider the conventional circular billiard in a 
magnetic field with strength $B$ directed perpendicular to the plane. The
classical motion of a particle with mass 
$m$, charge $q$, and speed $v$ confined inside a circle with specular
reflections is integrable; see Ref.~\cite{BB97} and references therein. 
The orbits consist of a series of arcs
of circles with the Larmor radius $R=mv/(qB$). 
Without loss of generality, we scale the radius of the billiard and the
absolute value of the momentum to unity. 
If $R<1$, some orbits form complete circles entirely inside the boundary.   
Ignoring these complete circles, we can specify each orbit by giving the
sequence of its positions and directions immediately after each impact at the
boundary. The position on the circular boundary is parametrized by the arclength $\phi \in [0,2\pi)$. The direction of the orbit after impact is
labeled by the angle of reflection $\alpha \in [0,\pi]$, or by the tangential
momentum $p = \cos{\alpha} \in [-1,1]$. Elementary geometry depicted in
\fig~\ref{fig:billiard} gives the Poincar\'e map, i.e., the discrete bounce
map from the $n$th to the $(n+1)$th collision with the boundary 
\bega\label{eq:circlemap1}
\phi_{n+1} & = & \phi_n + \omega(p_n)  \quad (\mbox{mod}\; 2\pi) \ ,\\
\label{eq:circlemap2}
p_{n+1} & = & p_n \ ,
\enda
with 
\bege\label{eq:omega}
\omega(p) = 2\arctan{\left(\frac{R\sqrt{1-p^2}}{1+Rp}\right)} \ ;
\ende
modulo $2\pi$ restricts the variable to the interval $[0,2\pi)$.
The function $\omega(p)$ is illustrated in \fig~\ref{fig:omega}. Note 
the broken time-reversal symmetry, $\omega(-p) \neq -\omega(p)$.   
\def\figbilliard{%
Part of a typical trajectory (solid arcs of circles) with specular (S) and
Andreev (A) reflections in the circular Andreev billiard. Dashed lines serve
for the construction of the maps~(\ref{eq:circlemap1})--(\ref{eq:Phi}).} 
\def\FIGbilliard{\centerline{\psfig{figure=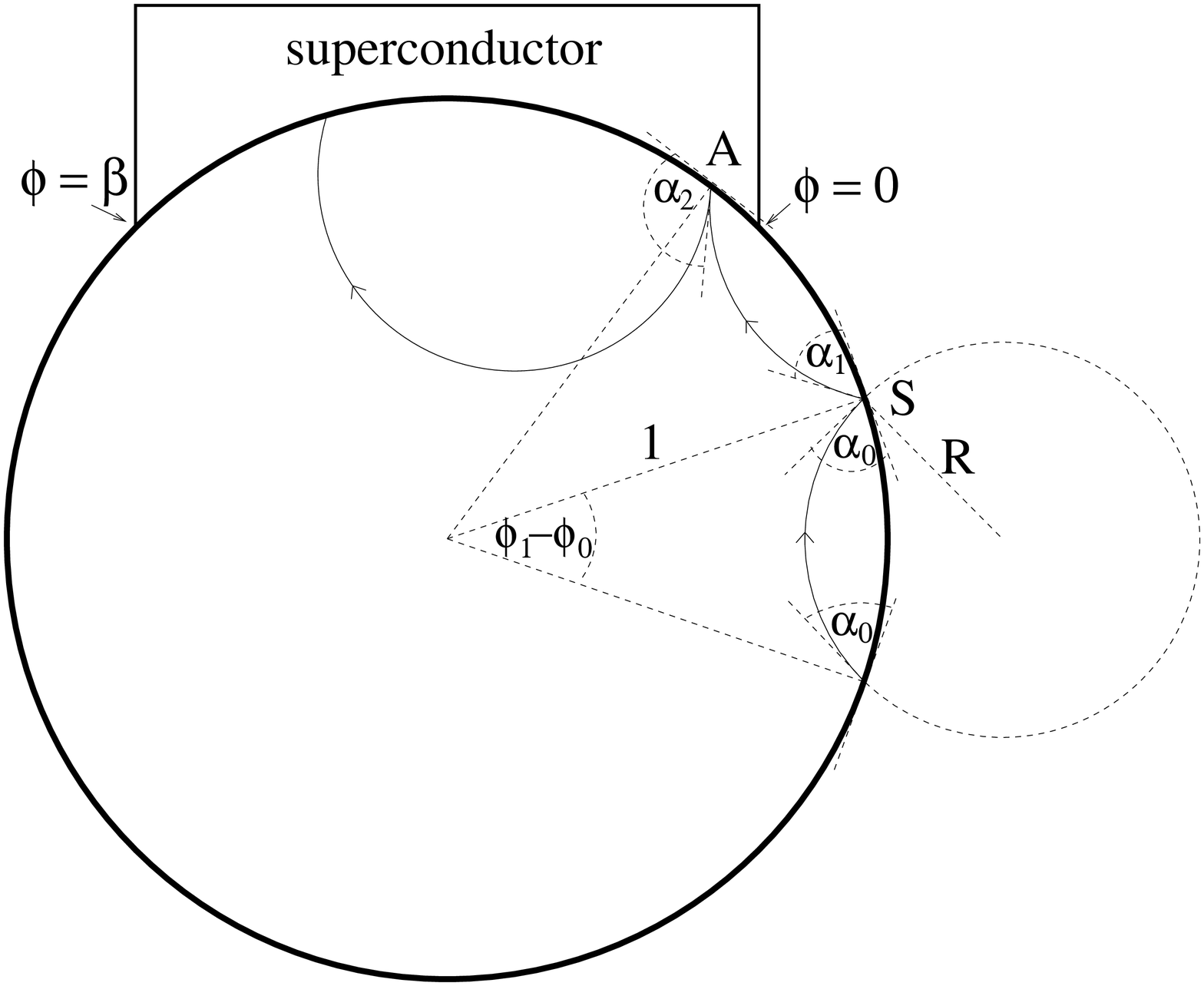,width=8.0cm,angle=0}
\vspace{0.2cm}
}}
\FIGo{fig:billiard}{\figbilliard}{\FIGbilliard}
\def\figomega{%
$\omega$ as function of $p$ according to \equ~(\ref{eq:omega}) with $R > 1$
(solid), $R = 1$ (dashed), and $R < 1$ (dotted). The lines $\omega = \pi$ and
$\omega = -\pi$ are identified.}
\def\FIGomega{\centerline{\psfig{figure=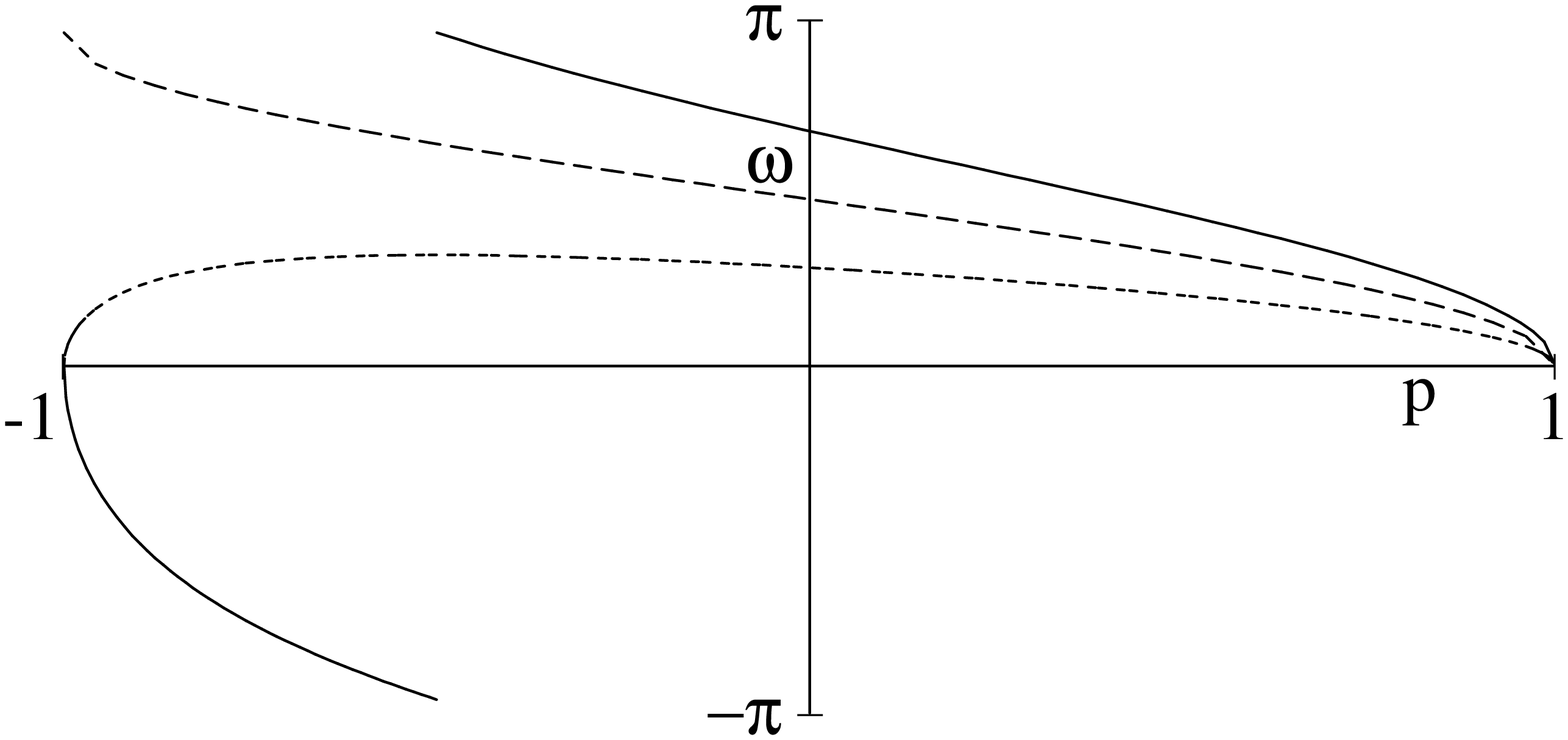,width=8.0cm,angle=0}
\vspace{0.2cm}
}}
\FIGo{fig:omega}{\figomega}{\FIGomega}

The properties of the map~(\ref{eq:circlemap1})--(\ref{eq:circlemap2}) are
related to the continuous-time evolution in a simple manner:
conservation of $p$ corresponds to conservation of angular momentum;
ergodic motion on an invariant circle $p = \const$ for irrational $\omega$ is
related to ergodic motion on a two-dimensional invariant torus; families of
fixed points for rational $\omega$ correspond to resonant tori foliated by
periodic orbits.   

The situation with a superconducting interface at $\phi \in (0,\beta)$ is
illustrated in \fig~\ref{fig:billiard}. 
For simplicity, let us first assume that the quasiparticles are exact at the
Fermi energy, i.e. $\varepsilon = 0$. 
An Andreev reflection at the interface is then just a change of sign of the
tangential momentum $p$; the replacement of an electronlike quasiparticle by a
holelike quasiparticle with the same energy and {\it vice versa} can be
ignored since the simultaneous change of the sign of the charge and the mass
does not alter the dynamics. 
Incorporating the change of the sign of $p$ at the interface into
the Poincar\'e map~(\ref{eq:circlemap1})--(\ref{eq:circlemap2}) gives 
\bega\label{eq:andreevmap1}
\phi_{n+1} & = & \phi_{n} + \omega(p_n)  \quad (\mbox{mod}\; 2\pi) \ ,\\
\label{eq:andreevmap2}
p_{n+1} & = & p_n \Phi(\phi_n)
\enda
with
\begin{equation}\label{eq:Phi}
\Phi(\phi) = \Biggl\{\begin{array}{cl}  
-1 & \mbox{if} \quad 0 < \phi < \beta \\ 
1 & \mbox{otherwise.} \end{array}
\end{equation}
The tangential momentum $p$ is no longer a constant of motion, but $|p|$ 
is. An invariant set $|p| = p_0 > 0$ consists of two circles $p =
p_0$ and $p = -p_0$, which are separated in phase space $(\phi,p)$. The
topology of this situation is schematically illustrated in
\fig~\ref{fig:dtorus}a.   
An invariant surface of the continuous-time dynamics is obtained from the
invariant set by attaching circles as shown in the sequences of
\figs~\ref{fig:dtorus}a-\ref{fig:dtorus}d.   
The circles represent the radial motion which is not contained in the
Poincar\'e map: after leaving the boundary, the radial coordinate decreases
until it reaches its minimum; then it increases again until it reaches its
maximum at the boundary. 
At a normal-conducting boundary point, this path is a full loop. 
Hence, we attach such a circle to each point of the set $\phi \in (\beta,2\pi)$
and $p \in \{p_0,-p_0\}$ (solid lines in \fig~\ref{fig:dtorus}b). 
At a superconducting boundary point, $p$ changes to $-p$ and we have
to trace the path once again to obtain a full loop.  
Hence, we connect each point $\phi \in (0,\beta)$ and $p = p_0$ with the
opposite point $p = -p_0$ (dashed
lines in \fig~\ref{fig:dtorus}c) and {\it vice versa}.
The resulting surface has the topology of a two-handled sphere (genus
2) as illustrated in \fig~\ref{fig:dtorus}d.    
At the critical points $\phi = 0$ and $\phi = \beta$ the motion is not well
defined. Two neighboring orbits hitting the boundary of the billiard at
different sides of a critical point separate from each other by moving along
different handles of the invariant surface. This is fully analogous to the
situation near critical corners in rational polygonal billiards; cf., e.g.,
\cite{RichensBerry81}. 
\def\figdtorus{
Construction of the invariant surfaces from the invariant sets.
Dotted lines mark the superconducting interface.
Thick dots mark the critical points.}
\def\FIGdtorus{\centerline{\psfig{figure=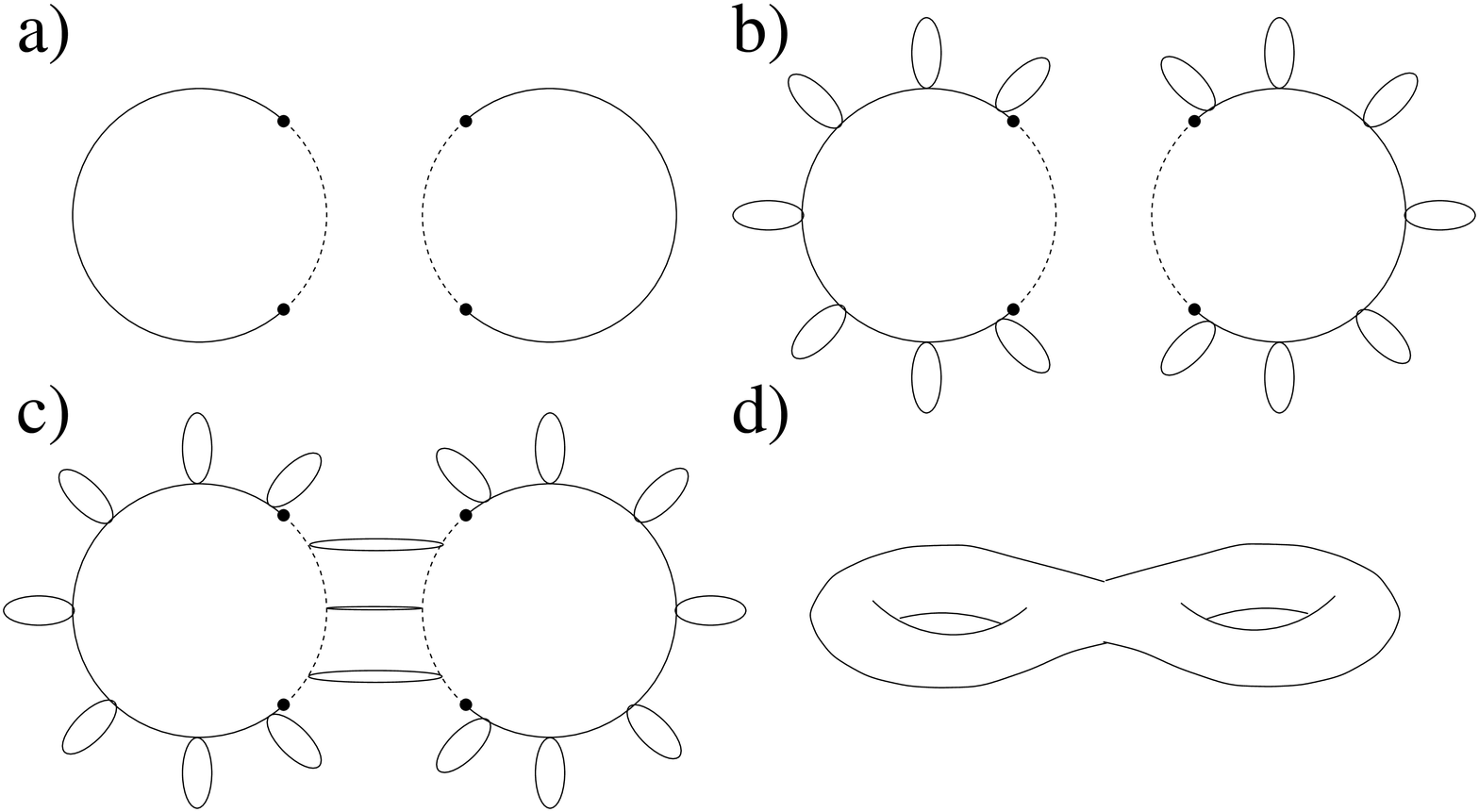,width=8.5cm,angle=0}
\vspace{0.25cm}
}}
\FIGo{fig:dtorus}{\figdtorus}{\FIGdtorus}

Let us now briefly demonstrate that the situation does not change qualitatively
when the quasiparticles are not exact at the Fermi energy, i.e. $\varepsilon >
0$.   
The electronlike quasiparticle with tangential momentum $p_e$ is
reflected in a holelike quasiparticle with $p_h \neq -
p_e$. Both particles have different Larmor radius, resulting in a
different $\omega$. However, if we redefine the tangential momentum as
$\tilde{p}= a p + b$ with $a=2c/(p_e-p_h)$, $b =
-a(p_e+p_h)/2$, $0 < c \leq 1$, and also redefine 
$\omega$ correspondingly then we recover the
map~(\ref{eq:andreevmap1})--(\ref{eq:andreevmap2}). Hence, it is sufficient to
consider $\varepsilon = 0$ as we will do in the following.

%%%%%%%%%%%%%%%%%%%%%%%%%%%%%%%%%%%%%%%
% barrier billiard
%%%%%%%%%%%%%%%%%%%%%%%%%%%%%%%%%%%%%%%
\section{The asymmetric barrier billiard}
\label{sec:barrier}
Before analysing the map~(\ref{eq:andreevmap1})--(\ref{eq:andreevmap2}) in
more detail, we discuss an interesting relation to a specific
rational polygonal billiard. A particle with unit mass
moves freely inside a polygon consisting of a vertical line of length $b$
placed in a rectangle with width $L^-+L^+$ and normalized height 1;
see~\fig~\ref{fig:barrier}. 
The symmetric case $L^-=L^+$ with $b = 1/2$ is the usual
barrier billiard~\cite{Zwanzig83,HM90,Wiersig01}; more general cases have been
considered in~\cite{Wiersig00,ZE01,EMS01}. 
Again, the trivial energy dependence is scaled away by setting the
energy to 1/2, or equally the magnitude of the momentum $(p_x,p_y)$ to
1. Starting with an initial momentum, only a finite number of directions can
be achieved during time evolution, stemming from the fact that all angles in
the polygon are rational multiples of $\pi$. In phase space, the motion takes
place on two-dimensional invariant surfaces $(|p_x|,|p_y|) = \const$. 
The general formula for the genus of such surfaces~\cite{RichensBerry81} gives
two due to the critical corner at the end of the barrier. 
\def\figbarrier{%
Asymmetric barrier billiard, rectangle with a vertical line connecting the
origin of the coordinate system $(x,y) = (0,0)$ with the point $(x,y) =
(0,b)$.}
\def\FIGbarrier{\centerline{\psfig{figure=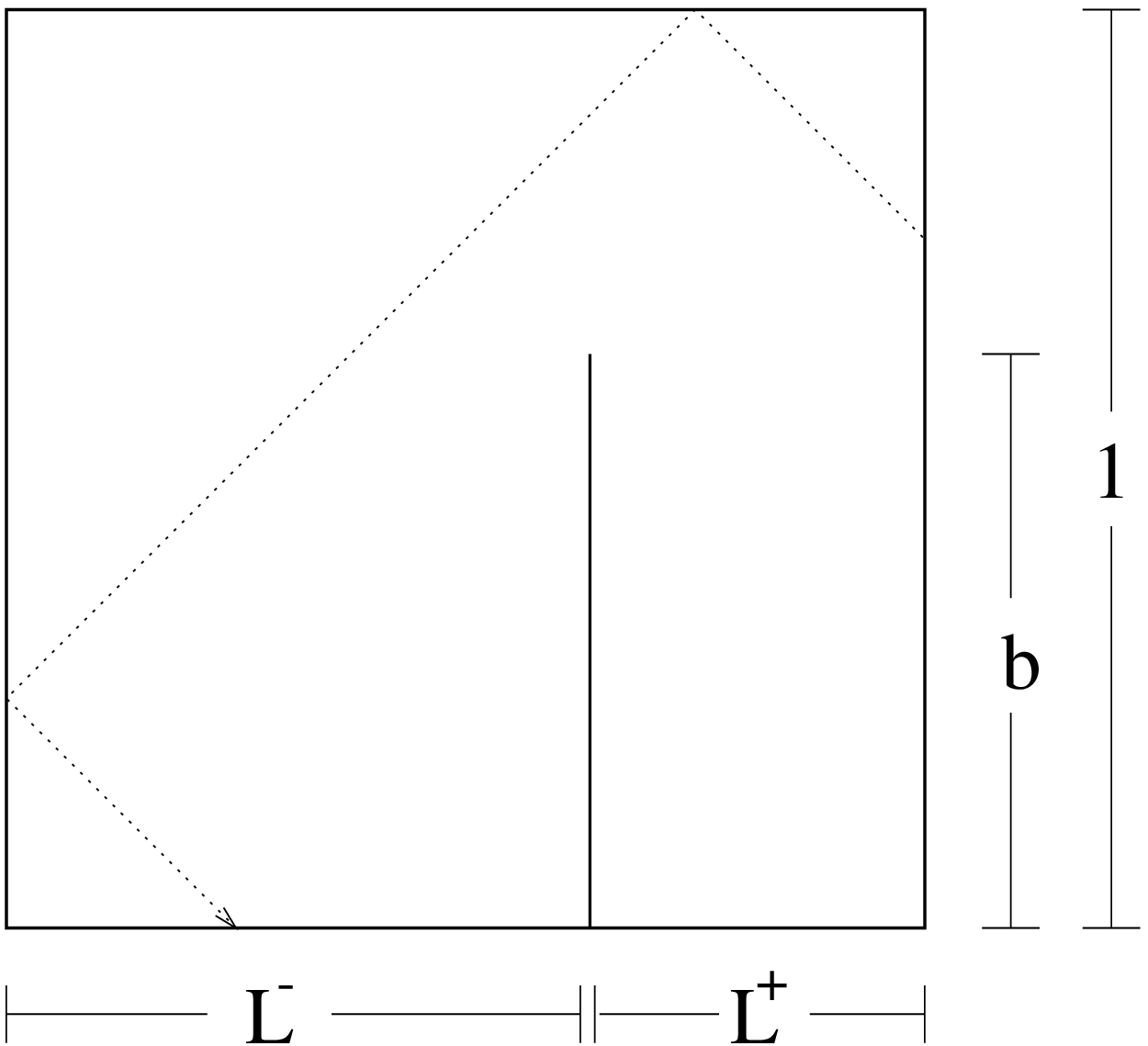,width=4.5cm,angle=0}
\vspace{0.25cm}
}}
\FIGo{fig:barrier}{\figbarrier}{\FIGbarrier}

It is convenient to consider the barrier billiard as two
rectangular billiards, one with $x \geq 0$ and one with $x\leq 0$, connected by
the passage $x=0$, $y > b$. Suppose for a moment that the passage
is closed, $b = 1$. For each of the two integrable rectangular
billiards we introduce action variables $I_x = |p_x|L/\pi$ and $I_y =
|p_y|/\pi$ where $L$ stands for $L^-$ and $L^+$. 
The time dependence of the angle variables is given by $\phi_x(t) = \omega_x t
\; (\mbox{mod}\; 2\pi)$ and $\phi_y(t) = \omega_y t \; (\mbox{mod}\; 2\pi)$  
with the frequencies $\omega_x = \pi^2I_x/L^2$ and $\omega_y = \pi^2I_y$.
The flow on the torus is ergodic if and only if the winding number
$\rho = \omega_y/\omega_x = I_y L^2/I_x$ is irrational. 
When the passage is open, $b < 1$, the particle can move from one
rectangle to the other one. We label the rectangles with the sign
of $x$, $s = \pm 1$. 
We then introduce the Poincar\'e section $x = 0$, i.e. we look at the line
in configuration space where $s(t)$ possibly changes.  
Choosing the origin of the angle variables and the barrier length $b$ such
that the passage is given by 
$0 < \phi_y < \beta = 2\pi(1-b)$ we get the map 
\bega\label{eq:barriermap1}
\phi_{y,n+1} & = & \phi_{y,n} + 2\pi\rho(s_n)  \quad (\mbox{mod}\; 2\pi) \ ,\\
\label{eq:barriermap2}
s_{n+1} & = & s_n \Phi(\phi_n) \ .
\enda

Even though the functions $2\pi\rho$ and $\omega$ are quite different,
the map~(\ref{eq:barriermap1})--(\ref{eq:barriermap2}) is related 
to the map~(\ref{eq:andreevmap1})--(\ref{eq:andreevmap2}) in the following
sense. Consider an orbit $(\phi_0,p_0),(\phi_1,p_1),\ldots$ in the Poincar\'e
map of the Andreev billiard~(\ref{eq:andreevmap1})--(\ref{eq:andreevmap2}). 
Denote the two ``frequencies'' as $\omega^+ = \omega(|p|)\;\mbox{mod}\;
2\pi$ and $\omega^- = \omega(-|p|)\;\mbox{mod}\;2\pi$.
The family of orbits parametrized by all accessible initial conditions
$(\phi_0,p_0)$ (with fixed $\omega^+$, $\omega^-$) is an invariant set and
related to an invariant surface of the Andreev billiard as described in
Sec.~\ref{sec:andreev}. 
Consider now an orbit $(\phi_{y,0},s_0),(\phi_{y,1},s_1),\ldots$ in the
map~(\ref{eq:barriermap1})--(\ref{eq:barriermap2}) with  
$2\pi\rho(+1) = \omega^+$ and $2\pi\rho(-1) = \omega^-$. This can be achieved
by constructing a barrier billiard with $L^+ = \omega^+|p_x/p_y|/(2\pi)$ and
$L^- = \omega^-|p_x/p_y|/(2\pi)$ where $(p_x,p_y)$ is the initial momentum.  
The family of orbits parametrized by all accessible initial conditions
$(\phi_{y,0},s_0)$ (with fixed $(p_x,p_y)$, $L^+$, and $L^-$) corresponds to
an invariant surface of the barrier billiard.
The two considered families of orbits are identical if we identify $\phi_y$
with $\phi$, and $s$ with $\mbox{sign}(p)$.
This interesting relation between orbits indicates that invariant surfaces in
the circular Andreev billiard not only have the same topology as in the 
barrier billiard, but also the dynamics on these surfaces (restricted to the 
chosen Poincar\'e surfaces of section) have typically the same ergodic
properties. We will discuss this issue in detail in the following section.
It is to emphasize that for our purpose it is not relevant that this kind of
equivalence is not complete (we have ignored the trivial circles  
lying entirely inside the billiard boundary for $R<1$) and possibly not
bijective (i.e. given a family of orbits in the 
barrier billiard there may be no counterpart in the Andreev billiard). A
complete correspondence between barrier and circular Andreev billiard is not
expected since the former one is symmetric under time reversal whereas the
latter one is not.   

Note that both maps have four points at which the dynamics is
discontinuous, $(\phi_y,s) = (0,\pm 1)$ and $(\phi_y,s) = (\beta,\pm 1)$. 
In the Andreev billiard these points are two copies of the two critical
points; see \fig~\ref{fig:dtorus}a. In the barrier billiard, the points
are four copies of the critical corner. 

\section{Dynamics on the invariant sets}
\label{sec:dynamics}
We now discuss the dynamics of the piecewise linear, area-preserving 
map~(\ref{eq:andreevmap1})--(\ref{eq:andreevmap2}) on the invariant sets. 
Clearly, the dynamics is not chaotic, since all Lyapunov exponents are zero. 
However, weaker ergodic properties, such as, mixing, weak mixing, and ergodicity
may be present.
First, ergodicity on the invariant sets follows directly from the fact that 
an orbit of the
map~(\ref{eq:andreevmap1})--(\ref{eq:andreevmap2}) has a counterpart in the
map~(\ref{eq:barriermap1})--(\ref{eq:barriermap2}) which is typically ergodic 
because the flow on the invariant surfaces in rational polygons is ergodic and
not mixing~\cite{Gutkin96}. 
However, mixing behavior of the 
map~(\ref{eq:andreevmap1})--(\ref{eq:andreevmap2}) cannot be excluded with 
this reasoning.  
We do this numerically by iterating
\equs~(\ref{eq:andreevmap1})--(\ref{eq:andreevmap2}) and computing the
time-averaged autocorrelation function 
(AF)   
\bege\label{eq:af}
R_z(j) = \frac{\langle z_{n+j}z_n\rangle}{\langle z^2_n \rangle} 
\ende
where $z$ stands for $p-\langle p\rangle$ and $\phi-\langle\phi\rangle$,
respectively. 
We have found that the AF typically does not decay to
zero which excludes the mixing property. 
However, the integrated AF 
\begin{equation}\label{eq:intAF}
R_{z,{\text {int}}}(n) = \frac1n\sum_{j=0}^n |R_z(j)|^2 
\end{equation}
is found to vanish according to a power law, $n^{-D_2}$ with $0< D_2 < 1$ for
large $n$; $D_2$ is the correlation dimension of the spectral 
measure~\cite{KPG92}.   
The example in \fig~\ref{fig:cor} shows that the integrated AF for both $p_n$
and $\phi_n$ clearly obeys the power law with $D_2 \approx 0.513$ and $D_2
\approx 0.577$, respectively.  
We observe qualitatively identical behavior for several nontrivial functions
$f(p_n,\phi_n)$ in a more or less pronounced way which is consistent with weak
mixing as maximal ergodic property.

However, this does not directly imply that the continuous-time evolution is
also weak mixing on the invariant surfaces. For example, the
barrier billiard in \fig~\ref{fig:barrier} is not weak mixing since $y(t)$ is
a periodic function.   
Nevertheless, since we cannot find such a trivial component in the
continuous-time evolution of the Andreev billiard, we believe that it is weak
mixing, but this certainly needs further studies. 
\def\figcor{%
Integrated autocorrelation function $R_{z,{\text {int}}}(n)$ for $\beta =
1/2$, $R = 1/4$, and $p_0 = 3/4$ in a ln-ln plot:
the upper set of points refers to $z = p-\langle p\rangle$ (solid line is the
linear fit), while the lower set of points refers to
$z = \phi-\langle\phi\rangle$ (dashed line).  
}
\def\FIGcor{\centerline{\psfig{figure=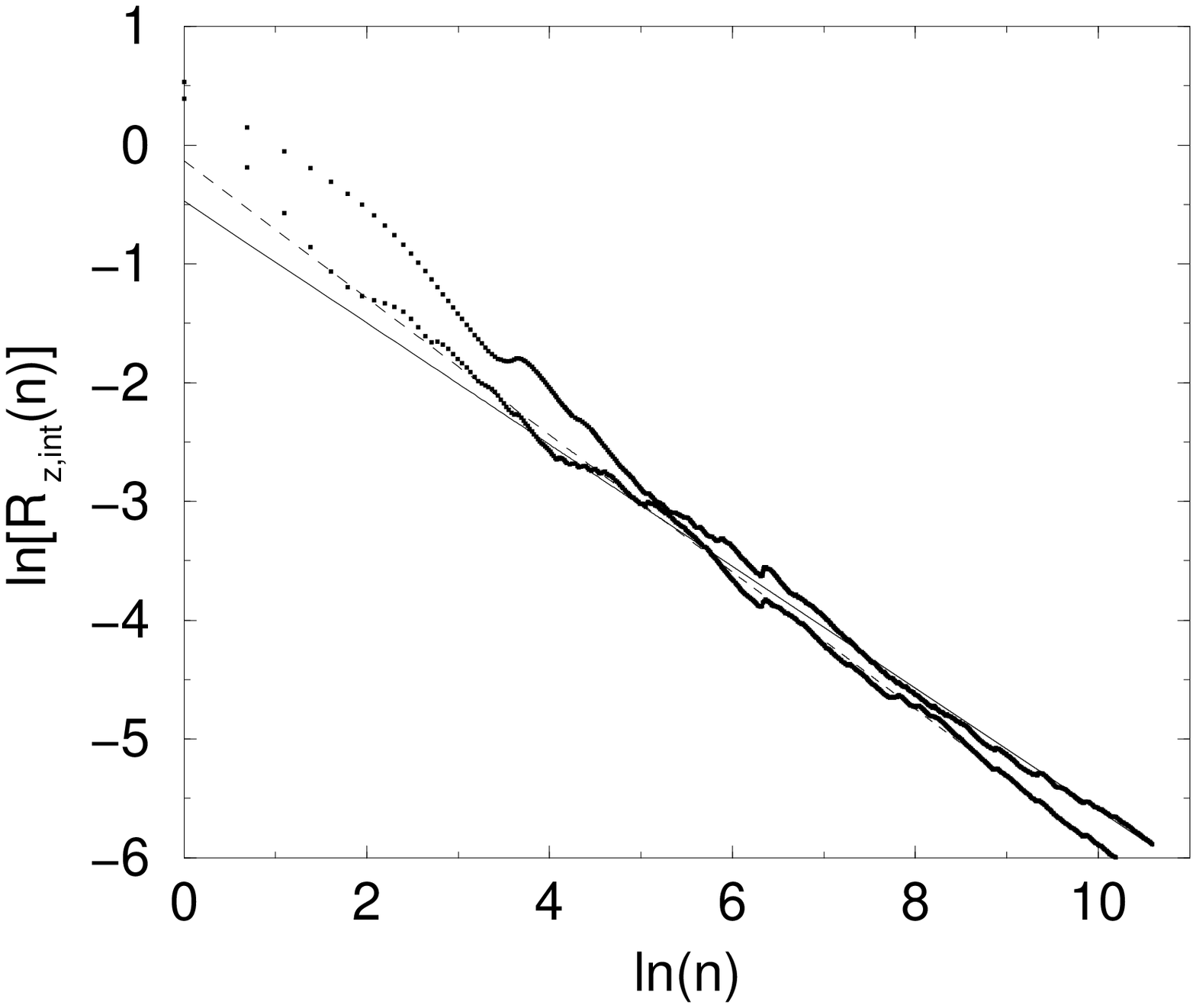,width=7.5cm,angle=0}
}}
\FIGo{fig:cor}{\figcor}{\FIGcor}

From our results we draw the following picture of the motion in
configuration space; cf. \fig~\ref{fig:billiard}. Starting with a small
cluster of initial conditions with fixed tangential momentum at the boundary
away from the superconducting interface, the particles move collectively along
skipping trajectories with constant mean angular velocity ($v\omega(p)$
divided by the arclength of the orbits) until the
superconducting region is met. Then, the sign of the tangential momentum is
inverted. Elementary geometry shows that the mean angular 
velocity changes. At the
next collision with the superconducting region the same thing happens. 
Orbits reaching the boundary at different sides of a critical point separate
after leaving the boundary. Hence, the cluster of initial conditions starts to 
spread out on the invariant surface. However, weak mixing allows
occasional reclusterings with decreasing frequency (the AF does not decay to
zero, but the integrated AF does). 

Finally, we address the two limiting cases of zero and high magnetic fields.
In the zero-field limit, time-reversal symmetry is recovered, $\omega(-p) =
-\omega(p)$. All orbits are self-retracing and periodic. The phase
space is foliated by periodic orbits rather than by two-dimensional surfaces.
In the regime of high magnetic field, $\omega(p)$ is small and close to
$\omega(-p)$. This situation corresponds to the symmetric barrier billiard
with small $\rho$.

\section{conclusion and outlook}
\label{sec:con}
We have studied a new kind of pseudointegrable system, the circular Andreev
billiard. It is different from known pseudointegrable systems in three
respects:  
(i) it does not belong to the class of polygonal billiards; 
(ii) it is not symmetric under time reversal; 
and (iii) it has a nontrivial foliation of energy surfaces by invariant
surfaces of genus two and two-parameter families of
periodic orbits (not touching the billiard boundary).
We have shown that the critical points, i.e., the boundary points separating
normal and superconducting regions, play the role of critical corners in 
polygons. 
Moreover, we have demonstrated that the dynamics on invariant surfaces
in the circular Andreev billiard have typically the same ergodic
properties as in the asymmetric barrier billiard (restricted to the chosen
Poincar\'e surfaces of section). 
This finding has been used to show that the Poincar\'e map is ergodic on the
invariant sets. Moreover, we have provided numerical evidence that the
Poincar\'e map is generically weak mixing. 

Weak mixing as maximal ergodic property implies interesting spectral
properties~\cite{CFS82} and anomalous transport. Recent studies in these
directions on rational polygons~\cite{ACG97,AGR00,Wiersig00} should be
easy to carry over to the circular Andreev billiard.

Of particular interest is the quantum mechanics of the circular Andreev
billiard because of two reasons.   
First, due to the pseudointegrability we expect an exotic quantum-classical
correspondence as in rational polygons~\cite{Wiersig01}.   
Second, as for rational polygons we might observe intermediate energy-level
statistics~\cite{BGS99} which are, however, modified by the broken 
time-reversal symmetry and the nontrivial foliation.    

The link between the two different classes of systems may also throw some light
on one aspect of the proximity effect in chaotic Andreev billiards, viz., the
appearance of a gap in the local density of states in an energy interval above
the Fermi energy.  
Random matrix theory can model this gap~\cite{MBFB96,MBFB97}, but the
semiclassical theory predicts an exponential 
suppression of the density of states~\cite{LN98,SB99}. 
One origin of this discrepancy could be the diagonal
approximation used in the semiclassical theory~\cite{ILVR01}. 
However, our finding indicates another possible explanation. It is well known
that the semiclassical treatment of polygonal billiards requires 
not only periodic orbits but also diffractive orbits, that are 
orbits starting and ending at critical corners; see e.g.,~\cite{BPS00} and
references therein.   
An interesting research project would be to incorporate analogously
diffractive orbits (stemming from the critical points) in the semiclassical
theory of Andreev billiards and see if this  
removes the discrepancy to the random matrix theory. 
This idea is supported by the fact that pointlike scatterers (diffraction) in
Andreev billiards gives rise to a gap in the spectrum near the Fermi
energy~\cite{PBB00}.

\section*{Acknowledgments}
I would like to thank H.~Schomerus for useful discussions and critically
reading the manuscript. 

% only for submission
%\newpage
\bibliographystyle{prsty}
\bibliography{../../bib/fg4,../../bib/extern}

%only for private preprint
\end{multicols}

% only for submission
\rem{
\newpage
\section*{Figure captions}
\setlength{\parindent}{0cm}
\printfigcap{fig:billiard}{\figbilliard}
\printfigcap{fig:omega}{\figomega}
\printfigcap{fig:dtorus}{\figdtorus}
\printfigcap{fig:barrier}{\figbarrier}
\printfigcap{fig:cor}{\figcor}

\newpage
\setcounter{figure}{0}
\showfig{\FIGbilliard}
\showfig{\FIGomega}
\showfig{\FIGdtorus}
\showfig{\FIGbarrier}
\showfig{\FIGcor}

%end rem
}

\end{document}